\documentclass{amsart}
\date{March 29, 2006}

\newtheorem{lemma}{Lemma}
\newtheorem{theorem}{Theorem}

\def\dbar{\mathrm{d}\mkern-6mu\mathchar'26} % dquer

\def\bp{\mathbf{p}}
\def\bq{\mathbf{q}}
\def\bx{\mathbf{x}}
\def\by{\mathbf{y}}
\def\bz{\mathbf{z}}
\def\bxi{\boldsymbol{\xi}}

\def\cE{\mathcal{E}}
\def\cF{\mathcal{F}}

\newcommand{\cz}{\mathbb{C}} % Komplexe Zahlen
\newcommand{\rz}{\mathbb{R}} % Relle Zahlen

\def\gH{\mathfrak{H}}
\def\gQ{\mathfrak{Q}}
\def\gS{\mathfrak{S}}
\def\gh{\mathfrak{h}}

\def\rd{\mathrm{d}}
\def\ri{\mathrm{i}}

\def\p{{\mathbf{\hat p}}} % Impulsoperator
\def\tr{\mathop\mathrm{tr}\nolimits} % trace

\title[Relativistic Energy]{The Ground State Energy of Heavy Atoms
  According to Brown and Ravenhall: Absence of Relativistic Effects in 
  Leading Order}

\author[R. Cassanas]{Roch Cassanas} 
\author[H. Siedentop]{Heinz Siedentop}

\address{Mathematisches Institut\\ Ludwig-Maximilians-Universit\"at
  M\"unchen\\ Theresienstra\ss e 39\\
  80333 M\"unchen\\
  Germany}

 \email{h.s@lmu.de \textrm{and} cassanas@math.lmu.de}



\thanks{This work was partially supported by the
    European Union's IHP network ``Analysis $\&$ Quantum''
    HPRN-CT-2002-00277.}

\begin{document}

\begin{abstract}  
  It is shown that the ground state energy of heavy atoms is, to
  leading order, given by the non-relativistic Thomas-Fermi energy.
  The proof is based on the relativistic Hamiltonian of Brown and
  Ravenhall which is derived from quantum electrodynamics yielding
  energy levels correctly up to order $\alpha^2$Ry.
\end{abstract}
\maketitle
\section{Introduction}
\label{sec1}
The energy of heavy atoms has attracted considerable interest in the
context of nonrelativistic quantum mechanics. Lieb and Simon
\cite{LiebSimon1977} proved that the leading behavior of the ground
state energy is given by the Thomas-Fermi energy which decreases as
$Z^{7/3}$. The leading correction to this behavior, the so called
Scott correction was established by Hughes
\cite{Hughes1986,Hughes1990} (lower bound), and Siedentop and Weikard
\cite{SiedentopWeikard1987O,SiedentopWeikard1987U,SiedentopWeikard1988,SiedentopWeikard1989,SiedentopWeikard1991}
(lower and upper bound). In fact even the existence of the
$Z^{5/3}$-correction conjectured by Schwinger was proven (Fefferman
and Seco
\cite{FeffermanSeco1989,FeffermanSeco1990,FeffermanSeco1990O,FeffermanSeco1992,FeffermanSeco1993,FeffermanSeco1994,FeffermanSeco1994T,FeffermanSeco1994Th,FeffermanSeco1995}).
Later these results where extended in various ways, e.g., to ions and
molecules.

Nevertheless, from a physical point of view, these considerations are
questionable, since large atoms force the innermost electrons on
orbits that are close to the nucleus where the electrons move with
high speed which requires a relativistic treatment. Our main goal in
this paper is to show that the leading energy contribution is
unaffected by relativistic effects, i.e., the asymptotic results of
Lieb and Simon \cite{LiebSimon1977} remain also valid in the
relativistic context, whereas the question mark behind the
quantitative correctness of the other corrections persists.

S\o rensen \cite{Sorensen2005} took a first step in this direction. He
considered the Chandrasekhar multi-particle operator and showed that
the leading energy behavior is given by the non-relativistic
Thomas-Fermi energy in the limit of large $Z$ and large velocity of
light $c$. Nevertheless, a question from the physical point of view
remains: Although the Chandrasekhar model is believed to represent
some qualitative features of relativistic systems, there is no reason
to assume that it should give quantitative correct results.
Therefore, to obtain not only qualitatively correct results it is
interesting, in fact mandatory, to consider a Hamiltonian which -- as
the one by Brown and Ravenhall \cite{BrownRavenhall1951} -- is derived
from QED such that it yields the leading relativistic effects in a
quantitave correct manner.

\section{Definition of the Model}
\label{sec2}

Brown and Ravenhall \cite{BrownRavenhall1951} describe two
relativistic electrons interacting with an external potential. The
model has an obvious generalization to the $N$-electron case.  The
energy in the state $\psi$ is defined as
\begin{equation}
  \label{eq:e}
  \begin{split}
    \cE:\bigwedge_{\nu=1}^N(H^{1/2}(\rz^3)\otimes\cz^4)&\rightarrow \rz\\
    \psi& \mapsto (\psi, (\sum_{\nu=1}^N(D_{c,Z} -c^2)_\nu+
      \sum_{1\leq\mu<\nu\leq N} |\bx_\mu-\bx_\nu|^{-1})\psi)
  \end{split}
\end{equation}
where 
$$D_{c,Z}:= \boldsymbol{\alpha}\cdot \frac c\ri\nabla + c^2\beta -Z
|\cdot|^{-1}$$
is the Dirac operator of an electron in the field of a
nucleus of charge $Z$.  As usual, the four matrices
$\alpha_1,...,\alpha_3$ and $\beta$ are the four Dirac matrices in
standard representation. We are interested in the restriction $\cE$ of
this functional onto
$\gQ_N:=\bigwedge_{\nu=1}^N(H^{1/2}(\rz^3)\otimes\cz^4)\cap \gH_N$
where
\begin{equation}
  \label{eq:2}
  \gH_N:=\bigwedge_{\nu=1}^N\gH;
\end{equation}
the underlying one-particle Hilbert space is
\begin{equation}
  \label{eq:1}
  \gH := [\chi_{(0,\infty)}(D_{c,0})](L^2(\rz^3)\otimes\cz^4).
\end{equation}

Note that we are using atomic units in this paper, i.e., $m_e=\hbar=e=1$.

As an immediate consequence of the work of Evans et al.
\cite{Evansetal1996} this form is bounded from below, in fact it is
positive (Tix \cite{Tix1997,Tix1998}), if $\kappa:=Z/c\leq
\kappa_\mathrm{crit}:=2/(\pi/2+2/\pi)$. (In the following, we will
assume that the ratio $\kappa\in[0,\kappa_\mathrm{crit})$ is fixed.)
According to Friedrichs this allows us to define a self-adjoint
operator $B_{c,N,Z}$ whose ground state energy
\begin{equation}
  \label{eq:5}  
  E(c,N,Z):= \inf\sigma(B_{c,N,Z})=\inf\{\cE(\psi)|\psi\in\gQ_N,
  \|\psi\|=1\}
\end{equation} 
is of concern to us in this paper. In fact -- denoting by
$E_\mathrm{TF}(Z,Z)$ the Thomas-Fermi energy of $Z$ electrons in the
field of nucleus with atomic number $Z$ and $q=2$ spin states per
electron (see Equations \eqref{eq:15} and \eqref{eq:15a} for more
details) -- our main result is
\begin{theorem}
  \label{t:haupt}
  $$E(Z/\kappa,Z,Z)=E_\mathrm{TF}(Z,Z)+o(Z^{7/3}).$$
\end{theorem}
This result, given here for the neutral atomic case, has obvious
generalizations to ions and molecules. To keep the presentation short
we refrain from presenting them here, as their treatment follows the
same strategy.

The remaining paper is structured as follows: First we show how the
treatment of the Brown-Ravenhall model can be reduced from Dirac
spinor (4-spinors) to Pauli spinors (2-spinors). In Section \ref{s2}
we prove the upper bound corresponding to Theorem \ref{t:haupt} by
rolling it back to Lieb's upper bound in the non-relativistic case
\cite{Lieb1981}. Section \ref{s3} reduces the lower bound to S\o
rensen's lower bound \cite{Sorensen2005}. Finally, in the appendix we
show that the correlation estimate using the exchange hole yields a
pointwise lower bound with uniform error of order $Z$. This is
interesting in itself since it allows to estimate the error purely by
the particle number not using any kinetic energy.

We now indicate, how to reduce to Pauli spinors. To this end we
parameterize the allowed states: Any $\psi\in \gH$ can be written as
\begin{equation}
  \label{eq:6}
  \psi :=
  \begin{pmatrix}
    {E_c(\p)+c^2\over N_c(\p)}u\\
    {c\p\cdot\boldsymbol\sigma\over N_c(\p)}u
  \end{pmatrix}
\end{equation}
for some $u\in \gh:=L^2(\rz^3)\otimes \cz^2$. Here,
$\boldsymbol{\sigma}$ are the three Pauli matrices,
$$\p:=-\ri\nabla,\ \ E_c(\bp):= (c^2\bp^2+c^4)^{1/2},\
N_c(\bp):=[2E_c(\bp)(E_c(\bp)+c^2)]^{1/2}.$$  In fact, the map
\begin{equation}
  \label{eq:7}
  \begin{split}
    \Phi: \gh &\rightarrow \gH\\
  u&\mapsto
  \begin{pmatrix}
    \Phi_1u\\
    \Phi_2u
  \end{pmatrix}
:=
\begin{pmatrix}
    {E_c(\p)+c^2\over N_c(\p)}u\\
    {c\p\cdot\boldsymbol\sigma\over N_c(\p)}u
  \end{pmatrix}
\end{split}
\end{equation}
embeds $\gh$ unitarily into $\gH$ and its restriction onto
$H^1(\rz^3)\otimes\cz^2$ is also unitary mapping to $\gH\cap
H^1(\rz^3)\otimes\cz^4$ (Evans et al. \cite{Evansetal1996}) ).

It suffices to study the
energy as function of $u$
\begin{equation}
  \label{eq:8}
  \cE\circ (\otimes_{\nu=1}^N\Phi): \bigwedge_{\nu=1}^N \gh \rightarrow \rz.
\end{equation}

The one-particle Brown-Ravenhall operator $B_\gamma$ for an electron
the external electric potential of a point nucleus acting on Pauli
spinors is then
\begin{equation}
  \label{eq:9}
  B_{c,Z}:= E_c(\p)- Z \varphi_1 -Z \varphi_2
\end{equation}
where we have split the potential into
\begin{equation}
  \label{eq:10}
  \varphi_1:=\Phi_1^*|\cdot|^{-1}\Phi_1,\ \
  \varphi_2:=\Phi_2^*|\cdot|^{-1}\Phi_2.
\end{equation}
As we will see the first part $\varphi_1$ is contributing to the
nonrelativistic limit whereas the second part turns out to give energy
contribution that do not even affect the first correction term.

\section{Upper Bound\label{s2}}
\subsection{Coherent States\label{ss2.1}}

The upper bound will be given by choosing a trial density matrix in the
Hartree-Fock functional for the Brown-Ravenhall operator. To this end
we introduce spinor valued coherent states.

Given any function $f\in H^{3/2}(\rz^3)$ and an element
$\alpha=(\bp,\bq,\tau)$ of the phase space
$\Gamma:=\rz^3\times\rz^3\times\{1,2\}$, we define coherent states in
$\gh$ as
\begin{equation}
  \label{eq:11}
  F_\alpha(x):= (\varphi_{\bp,\bq}\otimes e_\tau)(x)
  :=f(\bx-\bq)\exp(i\bp\cdot\bx)\delta_{\tau,\sigma},
\end{equation}
where $x=(\bx,\sigma)\in\rz^3\times\{1,2\}$ and the vectors $e_\tau$
are the canonical basis vectors in $\cz^2$ (see Lieb \cite{Lieb1981}
and Evans et al. \cite{Evansetal1996}). We will pick $f$ depending on
a dilation parameter. More specifically we will choose
\begin{equation}
  \label{eq:11a}
  f(\bx):=g_R(\bx):=R^{-3/2}g(R^{-1}\bx)
\end{equation}
where $R:=Z^{-\delta}$ with $\delta\in(1/3,2/3)$ and $g\in H^{3/2}$,
spherically symmetric, normalized, and with support in the unit ball.

The natural measure on $\Gamma$ counting the number of electrons per
phase space volume in the spirit of Planck is $\int_\Gamma
\dbar\Omega(\alpha) := (2\pi)^{-3}\int \rd \bp \int \rd\bq
\sum_{\tau=1}^2$. The essential properties needed are the following.
For $A\in L^1(\Gamma,\dbar \Omega)$ 
\begin{equation}
  \label{eq:11b}
  \gamma:= \int_\Gamma\dbar \Omega(\alpha) 
  A(\alpha)|F_\alpha\rangle\langle F_\alpha|
\end{equation} 
is a trace class operator and
\begin{eqnarray}
  \label{eq:12}
  0\leq A\leq1 \implies 0\leq \gamma\leq 1\\
\tr\gamma = \int_\Gamma \dbar \Omega(\alpha) A(\alpha). \label{eq:13}
\end{eqnarray}
Using $\Phi$ we can lift any such operator $\gamma$ to an operator on $\gH$
\begin{equation}
  \label{eq:14}
  \gamma_\Phi := \Phi\gamma\Phi^*.
\end{equation}

We will pick
\begin{equation}
  \label{eq:14a}
  A(\alpha):=\chi_{\{(\boldsymbol{\xi},\bx)\in\rz^6|\boldsymbol{\xi}^2/2 -
  V_Z(\bx)\leq 0\}}(\bp,\bq)
\end{equation}
where $V_Z:=Z/|\cdot|-|\cdot|^{-1}*\rho_\mathrm{TF}$; here
$\rho_\mathrm{TF}$ is the unique minimizer of the Thomas-Fermi
functional
\begin{equation}
  \label{eq:15}
  \cE_\mathrm{TF}(\rho):= \int_{\rz^3}\left[\frac35\gamma_\mathrm{TF}\rho(\bx)^{5/3} - \frac Z{|\bx|} \rho(\bx)\right]\rd \bx + D(\rho,\rho)
\end{equation}
where, for Fermions with $q$ spin states per particle,
$\gamma_\mathrm{TF}:=(6\pi^2/q)^{2/3}\hbar^2/(2m)$, i.e., in our
units, $\gamma_\mathrm{TF}=(3\pi^2)^{2/3}/2$. Note that
$\int\dbar\Omega(\alpha) A(\alpha)=Z$ (Lieb and Simon
\cite{LiebSimon1977}). Note also that $V_z(\bq):=
Z^{4/3}V_1(Z^{1/3}\bq)$ (see also Gomb\'as \cite{Gombas1949} and
\cite{LiebSimon1977}). Note also that the minimal energy
$E_\mathrm{TF}(N,Z)$ fulfills the scaling relation
\begin{equation}
  \label{eq:15a}
  E_\mathrm{TF}(N,Z)=E_\mathrm{TF}(N/Z,1)Z^{7/3}.
\end{equation}

Note, that we could restrict the minimization to $\int \rho\leq N$
without any problem. For $N\geq Z$ there would be no change in the
minimizer; for $N < Z$ we would get a different minimizer. For
notational convenience we will merely consider the neutral case $N=Z$
in the following.

\subsection{Upper Bound\label{ss3.2}}

We begin by noting that the Hartree-Fock functional -- with or without
exchange energy -- bounds $E(c,N,Z)$ from above. To be exact we
introduce the set of density matrices
\begin{equation}
  \label{eq:16}
  S_N:=\{\gamma\in \gS^1(\gh)\ | E_c(\hat\bp)\gamma\in \gS^1(\gh),\ 0\leq\gamma\leq1,\ \tr\gamma=N\}
\end{equation}
where $\gS^1(\gh)$ denotes the trace class operators on $\gh$.
\begin{equation}
  \label{eq:17}
  \begin{split}
    \cE_\mathrm{HF}: S_N&\rightarrow \rz\\
    \gamma&\mapsto \tr[(E_c(\hat\bp)-c^2 - Z/|\bx|)\gamma_\Phi]+ D(\rho_{\gamma_\Phi},\rho_{\gamma_\Phi}) 
  \end{split}
\end{equation}
where -- as usual -- $\rho_\gamma$ is the density associated to
$\gamma$ and $D$ is the Coulomb scalar product. By the analogon of
Lieb's result \cite{Lieb1981V,Lieb1981E} (see also Bach
\cite{Bach1992}) -- which trivially transcribes from the Schr\"odinger
setting to the present one -- we have for all $\gamma\in S_N$
\begin{equation}
  \label{eq:18}
  E(c,N,Z)\leq \cE_\mathrm{HF}(\gamma).
\end{equation}

\subsubsection{Kinetic Energy\label{sss3.2.1}}

By concavity we have
\begin{equation}
  \label{eq:19}
  E_c(\bp)-c^2\leq \tfrac12\bp^2
\end{equation}
which implies that the Brown-Ravenhall kinetic energy is bounded by
the non-relativistic one, i.e., for all $\gamma \in S_N$ with
$-\Delta\gamma\in \gS^1(\gh)$
\begin{equation}
  \label{eq:20}
  \tr[(E_c(\hat\bp)-c^2)\gamma]\leq \tr(-\tfrac12\Delta\gamma).
\end{equation}
Inserting our choice of $\gamma$ (see Equations \eqref{eq:11}, \eqref{eq:11a},
\eqref{eq:11b}, and \eqref{eq:14a}) turns the right hand side into the
Thomas-Fermi kinetic energy modulo the positive error $Z \|\nabla
g\|^2 R^{-2}$ (see Lieb \cite[Formula (5.9)]{Lieb1981}), i.e.,
\begin{equation}
  \label{eq:21}
  \tr[(E_c(\hat\bp)-c^2)\gamma] 
  \leq \frac35\gamma_\mathrm{TF}\int \rho_\mathrm{TF}^{5/3}(\bx)\rd\bx 
  + ZR^{-2}\|\nabla g\|^2.
\end{equation}

\subsubsection{External Potential\label{sss3.2.2}}

Since $-Z\tr(\varphi_2\gamma)$ is negative, we can and will estimate
this term by zero. This estimate will be good, if this term is of
smaller order. Although, logically unnecessary for the upper bound, it
is, for pedagogical reasons, interesting to see that $\varphi_2$ does
indeed not significantly contribute to the energy, if $\gamma$ is
chosen as above. Moreover, the proof will be also useful for the proof of
Lemma \ref{l2}.
\begin{lemma}
  \label{l1}
  For our choice of
  $\gamma=\int_\Gamma\dbar\Omega(\alpha)|F_\alpha\rangle\langle
  F_\alpha|$ and $\delta\in(1/3,2/3)$ we have
  \begin{equation}
    \label{eq:22}
    0\leq Z\tr(\varphi_2\gamma) 
    \leq k Z\int_\Gamma\dbar\Omega(\alpha)A(\alpha)
    \iint\rd \bxi \rd \bxi'{c^2|\bxi||\bxi'||\hat F_\alpha(\bxi)||\hat F_\alpha(\bxi')| \over |\bxi-\bxi'|^2 N_c(\bxi)N_c(\bxi')} 
    = O(Z^{4/3+\delta}).
  \end{equation}
\end{lemma}
(In the following -- throughout the paper -- we use the letter $k$ for a
constant independent of $c$, $N$, $R$, or $Z$.)
\begin{proof}
  We begin by estimating the expectation of $\varphi_2$ in a coherent
  state. 
  \begin{multline}
    \label{eq:23}
    0 \leq (F_\alpha, \varphi_2 F_\alpha) \leq k \iint \rd \bxi \rd
    \bxi' {c^2|\bxi||\bxi'| |\hat F_\alpha(\bxi) \hat F_\alpha(\bxi')| \over
      N_c(\bxi)|\bxi-\bxi'|^2
      N_c(\bxi')} \\
    \leq k c^{-2}R^{-3} \iint \rd\bxi\rd\bxi'{|\hat g(\bxi)\hat
      g(\bxi')|\over |\bxi-\bxi'|^2}|\bxi+R\bp||\bxi'+R\bp| \leq k
    c^{-2}R^{-3} (1+ R|\bp|+R^2|\bp|^2).
  \end{multline}
  Here, we used that $N_c(\bxi)\geq\sqrt2c^2$ and, in the last step,
  that
  $$
  {|\hat g(\bxi)\hat g(\bxi')|\over |\bxi-\bxi'|^2}[|\bxi||\bxi'| +
  |\bxi|+|\bxi'| +1]
  $$
  is integrable in $\bxi$ and $\bxi'$ because $g\in H^{3/2}(\rz^3)$. Thus we get
  \begin{multline}
    \label{eq:24}
    0\leq Z\tr(\varphi_2\gamma)=Z\int \dbar\Omega(\alpha) A(\alpha)(F_\alpha, \varphi_2 F_\alpha)\\
    \leq k{Z\over c^2R^3} \int \dbar\Omega(\alpha) A(\alpha)(1+ R|\bp|+R^2|\bp|^2)\\
    \leq  k{Z\over c^2R^3}\left\{Z+ R\int\rd \bq \left[Z^{4/3} V_1(Z^{1/3}\bq)\right]^2+R^2 \left[Z^{4/3} V_1(Z^{1/3}\bq)\right]^{5/2}\right\}\\
    = O(Z^{3\delta} + Z^{2/3+2\delta} + Z^{4/3+\delta}).
  \end{multline}
\end{proof}
\begin{lemma}
  \label{l2}
  For our choice of $\gamma$ and $\delta\in(1/3,2/3)$ we have
  \begin{multline}
    \label{eq:25}
    \left|Z\tr[(|\cdot|^{-1}-\varphi_1)\gamma]\right| \\
    \leq k Z\int\dbar\Omega(\alpha)A(\alpha) \iint\frac{\rd\boldsymbol{\xi}\rd\boldsymbol{\xi}'}{|\boldsymbol{\xi}-\boldsymbol{\xi}'|^2}
    \left(1-{(E_c(\boldsymbol{\xi})+c^2)(E_c(\boldsymbol{\xi}')+c^2) \over
      N_c(\boldsymbol{\xi})N_c(\boldsymbol{\xi}')}\right) |\hat F_\alpha(\boldsymbol{\xi})||\hat F_\alpha(\boldsymbol{\xi}')| \\
    = O(Z^{5/3+\delta}).
  \end{multline}
\end{lemma}
\begin{proof}
  We fisrt note that
  \begin{multline}
    \label{eq:26a} 
    \left|1-{(E_c(\boldsymbol{\xi})+c^2)(E_c(\boldsymbol{\xi}')+c^2)
        \over N_c(\boldsymbol{\xi})N_c(\boldsymbol{\xi}')}\right| \\
    \leq
    {\left|3E_c(\boldsymbol{\xi})E_c(\boldsymbol{\xi}')-c^2(E_c(\boldsymbol{\xi})+E_c(\boldsymbol{\xi}))+c^2\right|
      \over N_c(\boldsymbol{\xi})N_c(\boldsymbol{\xi}')}.
  \end{multline}
  Then, noting that $E_c(\boldsymbol{\xi})-c^2\leq
  c|\boldsymbol{\xi}|$, we obtain
  \begin{multline}
  \label{eq26b}
  \left|1-{(E_c(\boldsymbol{\xi})+c^2)(E_c(\boldsymbol{\xi}')+c^2) \over
      N_c(\boldsymbol{\xi})N_c(\boldsymbol{\xi}')}\right|
  \leq {3c^2|\boldsymbol{\xi}||\boldsymbol{\xi}'|+2c^3(\boldsymbol{\xi}+\boldsymbol{\xi}')\over
    N_c(\boldsymbol{\xi})N_c(\boldsymbol{\xi}')}\\
  \leq{3c^2|\boldsymbol{\xi}||\boldsymbol{\xi}'|+2c^3(\boldsymbol{\xi}+\boldsymbol{\xi}')\over
      2c^4} .
  \end{multline}
  Using this last equation, we estimate
  \begin{multline}
    \label{eq:26}
    |(F_\alpha, (\frac1{|\cdot|}-\varphi_1) F_\alpha)|
    \\
    \leq k
    \iint\frac{\rd\boldsymbol{\xi}\rd\boldsymbol{\xi}'}{|\boldsymbol{\xi}-\boldsymbol{\xi}'|^2}
    \left(1-{(E_c(\boldsymbol{\xi})+c^2)(E_c(\boldsymbol{\xi}')+c^2)
        \over
        N_c(\boldsymbol{\xi})N_c(\boldsymbol{\xi}')}\right) |\hat F_\alpha(\boldsymbol{\xi})||\hat F_\alpha(\boldsymbol{\xi}')| \\
    \leq k \int_{\rz^6} \rd \boldsymbol{\xi}\rd\boldsymbol{\xi}'
    {|\hat g_R(\boldsymbol{\xi}-\bp) \hat g_R(\boldsymbol{\xi}'-\bp)|
      \over |\boldsymbol{\xi}-\boldsymbol{\xi}'|^2}   (c^{-2}|\boldsymbol{\xi}||\boldsymbol{\xi}'| +c^{-1} (|\boldsymbol{\xi}|+|\boldsymbol{\xi}'|))\\
    \leq k c^{-2}R^{-3} \int
    \rd\boldsymbol{\xi}\int\rd\boldsymbol{\xi}'{|\hat
      g(\boldsymbol{\xi})\hat
      g(\boldsymbol{\xi}')|\over |\boldsymbol{\xi}-\boldsymbol{\xi}'|^2}( |\boldsymbol{\xi}+R\bp||\boldsymbol{\xi}'+R\bp| + c R|\boldsymbol{\xi}+R\bp|)\\
    \leq k c^{-2}R^{-3} \int
    \rd\boldsymbol{\xi}\int\rd\boldsymbol{\xi}'{|\hat
      g(\boldsymbol{\xi})\hat g(\boldsymbol{\xi}')|\over
      |\boldsymbol{\xi}-\boldsymbol{\xi}'|^2}(
    |\boldsymbol{\xi}||\boldsymbol{\xi}'|+R|\bp||\boldsymbol{\xi}|+|R\bp|^2
    + cR|\boldsymbol{\xi}|+cR^2|\bp|)\\
    \leq k c^{-2}R^{-3} (1+ R|\bp|+R^2|\bp|^2+cR +cR^2|\bp|).
  \end{multline}
  Thus
  \begin{multline}
    \label{eq:27}
    Z|\tr[(|\cdot|^{-1}-\varphi_1)\gamma]\leq
    Z|\int_\Gamma\dbar\Omega(\alpha)
    A(\alpha)(F_\alpha, (|\cdot|^{-1}-\varphi_1) F_\alpha)|\\
    \leq k Z\int\dbar\Omega(\alpha)A(\alpha)
    \iint\frac{\rd\boldsymbol{\xi}\rd\boldsymbol{\xi}'}{|\boldsymbol{\xi}-\boldsymbol{\xi}'|^2}
    \left(1-{(E_c(\boldsymbol{\xi})+c^2)(E_c(\boldsymbol{\xi}')+c^2)
        \over
        N_c(\boldsymbol{\xi})N_c(\boldsymbol{\xi}')}\right) |\hat F_\alpha(\boldsymbol{\xi})||\hat F_\alpha(\boldsymbol{\xi}')| \\
    \leq k(Z^{3\delta}+ Z^{2\delta + 2/3}+ +Z^{\delta+4/3}+
    Z^{2\delta}+ Z^{\delta+5/3})
  \end{multline}
  which yields the desired estimate.
\end{proof}

\subsubsection{The Electron-Electron Interaction\label{sss3.2.3}}

We will roll back the treatment of the electron-electron interaction
to the treatment of nucleus-electron interaction.

\begin{lemma}
  \label{l3}
  For our choice of $\gamma$ and $\delta\in(1/3,2/3)$ we have
  \begin{equation}
    \label{eq:28}
    D(\rho_{\gamma_\Phi},\rho_{\gamma_\Phi})-D(\rho_\gamma,\rho_\gamma)= O(Z^{5/3+\delta}),
  \end{equation}
  where $\rho_\gamma$ is the density of $\gamma$ and $\rho_{\gamma_\Phi}$ is the
  density of $\gamma_\Phi$.
\end{lemma}
\begin{proof}
  We have
  \begin{equation}
    \label{eq:29}
    |\cF[(\rho_\gamma+\rho_{\gamma_\Phi})*|\cdot|^{-1}](\bxi)|\leq \sqrt{2/\pi}\|\rho_\gamma + \rho_{\gamma_\Phi}\|_1|\bxi|^{-2} =2^{3/2}\pi^{-1/2} Z|\bxi|^{-2}.
  \end{equation}
  Now,
\begin{multline*}
  |D(\rho_{\gamma_\Phi},\rho_{\gamma_\Phi})-D(\rho_\gamma,\rho_\gamma)|=
  |D(\rho_{\gamma_\Phi}-\rho_\gamma,\rho_{\gamma_\Phi}+\rho_\gamma)|\\
  \leq \frac12\left|\int_{\rz^3}(\rho_\gamma(\bx)-\rho_{\gamma_\Phi}(\bx))[(\rho_\gamma+\rho_{\gamma_\Phi})*|\cdot|^{-1}](\bx)\rd \bx\right|\\
  \leq \frac12\int_\Gamma \dbar\Omega(\alpha)A(\alpha)\\
  \times\iint \rd\bxi \rd\bxi'|
  \cF[(\rho_\gamma+\rho_{\gamma_\Phi})*|\cdot|^{-1}](\bxi-\bxi')|K(\bxi,\bxi')
  |\hat
  F_\alpha(\bxi)||\hat F_\alpha(\bxi')| \rd\bxi\rd\bxi'\\
  \leq \sqrt\frac2\pi Z\int_\Gamma \dbar\Omega(\alpha)A(\alpha)\iint \rd\bxi
  \rd\bxi'| |\bxi-\bxi'|^{-2} K(\bxi,\bxi') |\hat F_\alpha(\bxi)||\hat
  F_\alpha(\bxi')| \rd\bxi\rd\bxi'
\end{multline*}
where
$$K(\bxi,\bxi')=\left|\frac{(E_c(\bxi)+c^2)(E_c(\bxi')+c^2)}{N_c(\bxi)N_c(\bxi')}-1 \right| +  \frac{c^2|\bxi||\bxi'|}{N_c(\bxi)N_c(\bxi')}$$
and where we used \eqref{eq:29} in the last step. Eventually, applying
Lemmata \ref{l1} and \ref{l2} yields the desired result.
\end{proof}

\subsubsection{The Total Energy\label{sss3.2.4}}

Gathering our above estimates allows us to reduce the problem to the
non-relativistic result of Lieb \cite{Lieb1981}
\begin{theorem}
  We have $E(\kappa Z, Z,Z)\leq E_\mathrm{TF}(1,1)Z^{7/3} + k Z^{20/9}$.
\end{theorem}
\begin{proof}
  Following Lieb \cite[Section V.A.1]{Lieb1981} with the remainder
  terms given there (putting $R=Z^{-\delta}$ as in our estimate),
  using the remainder terms obtained in Lemmata \ref{l1} through
  \ref{l3}, and using \eqref{eq:21} we get
  \begin{equation}
    \label{eq:31}
    E(c,Z,Z)\leq \cE_\mathrm{HF}(\gamma) \leq E_\mathrm{TF}(Z,Z) +
    O(Z^{1+2\delta} + Z^{\frac52-\frac\delta2} + Z^{\frac53+\delta})
  \end{equation}
  which is optimized for $\delta=5/9$ giving the claimed result.
\end{proof}

\section{Lower Bound\label{s3}}

The lower bound is -- contrary to the usual folklore -- easy. As we
will see, it is a corollary of S\o rensen's \cite{Sorensen2005} result
for the Chandrasekhar operator and an estimate on the potential
generated by the exchange hole \cite{Mancasetal2004}.  The exchange
hole of a density $\sigma$ at a point $\bx\in\rz^3$ is defined as the
ball $B_{R_\sigma(\bx)}(\bx)$ of radius $R_\sigma(\bx)$ centered at
$\bx$ where $R_\sigma(\bx)$ is the smallest radius $R$ fulfilling
  \begin{equation}
    \label{eq:31a}
    \frac12=\int_{B_R} \sigma.
  \end{equation}
The hole potential $L_\sigma $ of $\sigma$ is defined through
\begin{equation}
  \label{eq:31b}
  L_\sigma(\bx):=  \int_{|\bx-\by|<R_\sigma(\bx)}{\sigma(\by) \over |\bx-\by| }\rd\by.
\end{equation}

  Our second main result is the following lower bound.
\begin{theorem}
  \label{t2}
  $$\liminf_{Z\to\infty}[(E(c,Z,Z)-E_\mathrm{TF}(Z,Z)]Z^{-7/3} \geq 0.$$
\end{theorem}
\begin{proof}
  Pick $\delta>0$ and set $\rho_\delta:=\rho_\mathrm{TF}*g_{Z^{-\delta}}^2.$ Then
  the exchange hole correlation bound \cite[Equation (14)]{Mancasetal2004} implies
  the following pointwise estimate
  \begin{equation}
    \label{eq:32}
    \sum_{1\leq\mu<\nu\leq N}{1\over|\bx_\mu-\bx_\nu|}\geq
    \sum_{\nu=1}^N[\rho_\delta*|\cdot|^{-1}(\bx_\nu)- L_{\rho_\delta}(\bx_\nu)]-
    D(\rho_\delta,\rho_\delta).
  \end{equation}
  Because of the spherical symmetry of $g$ we can use Newton's theorem
  \cite{Newton1972} and replace $\rho_\delta$ by $\rho_\mathrm{TF}$ in
  the third summand of the right hand side of \eqref{eq:32}. Then, by
  Lemma \ref{l8}, we get that for all normalized $\psi\in\gQ_N$
  \begin{equation}
    \label{eq:32a}
    \cE(\psi)    \geq \tr[\Lambda_+(|D_0|-c^2-V_\delta)\Lambda_+]_- 
    -kNZ -D(\rho_{\mathrm{TF}},\rho_{\mathrm{TF}})
  \end{equation}
  where, for $t\in\rz$, we set $[t]_-:=\min\{ t,0 \}$ and
  $V_\delta=Z/|\cdot|-\rho_\delta*|\cdot|^{-1}$.
  
  To count the number of spin states per electron correctly, i.e., two
  instead of the apparent four, we use an observation by Lieb et al.
  \cite[Appendix B]{Liebetal1997}: Note that
  \begin{equation}
    \label{eq:32b}
    \Lambda_-=U^{-1}\Lambda_+ \, U,\qquad \mbox{ where }\quad
    U:=\left(
      \begin{array}{cc}
        0    & 1\\
        -1 & 0
      \end{array}\right).
  \end{equation}
  Indeed, we have
  $$\Lambda_-=\frac12\left(1-\frac{D_0}{|D_0|}\right), \qquad
  \Lambda_+=\frac12\left(1+\frac{D_0}{|D_0|}\right)$$
  and
  $$U^{-1}D_0 \, U=
  \left(\begin{array}{cc}
      0    & 1\\
      -1 & 0
    \end{array}\right)
  \left(\begin{array}{cc}
      mc^2         & c\,\sigma.\hat{p}\\
      c\, \sigma.\hat{p} & -mc^2 
    \end{array}\right)
  \left(\begin{array}{cc}
      0   & -1\\
      1 & 0
    \end{array}\right)=-D_0.$$
  We set $X:=(|D_0|-c^2-V_\delta(x))I_2$, and write
  $$\tr\left[\Lambda_+\left(\begin{array}{cc}
        X   & 0\\
        0   & X
      \end{array}\right)\Lambda_+\right]_-\geq \tr\left( \Lambda_+
    \left(\begin{array}{cc}
        X_-   & 0\\
        0     & X_-
      \end{array}\right)\Lambda_+\right)=
  \tr\left( \Lambda_+
    \left(\begin{array}{cc}
        X_-   & 0\\
        0     & X_-
      \end{array}\right)\right)$$
  $$\tr\left( \Lambda_-
    \left(\begin{array}{cc}
        X_-   & 0\\
        0     & X_-
      \end{array}\right)\right)=
  \tr\left( \Lambda_+ U
    \left(\begin{array}{cc}
        X_-   & 0\\
        0     & X_-
      \end{array}\right)U\right)=
  \tr\left( \Lambda_+
    \left(\begin{array}{cc}
        X_-   & 0\\
        0     & X_-
      \end{array}\right)\right)$$
  Thus
  \begin{multline}
    \label{eq:32c}
    2\tr\left( \Lambda_+ \left(\begin{array}{cc}
          X_-   & 0\\
          0 & X_-
        \end{array}\right)\right)\\
    = \tr\left( \Lambda_+ \left(\begin{array}{cc}
          X_-   & 0\\
          0 & X_-
        \end{array}\right)\right)+
    \tr\left( \Lambda_-
      \left(\begin{array}{cc}
          X_-   & 0\\
          0     & X_-
        \end{array}\right)\right)
    =2\tr(X_-).
  \end{multline}
  Since $|D_0|=E_c(\hat{p})$, we obtain
  \begin{equation}
    \label{eq:32d}
    E(\kappa Z,Z,Z)\, \geq \,2\, \tr[E_c(\hat{p})-c^2-V_\delta(x)]_- 
    -D(\rho_{\mathrm{TF}},\rho_{\mathrm{TF}})-kNZ
  \end{equation}
  where the last trace is spinless. This connects to S\o rensen's
  Equation (3.2) from \cite{Sorensen2005}. The result then follows
  using his lower bound.
\end{proof}

\appendix
\section{$L^\infty$-Bound on the Exchange Hole Potential\label{a1}}

We begin the appendix with the following remark: the Thomas-Fermi
potential $V_Z:=Z/|\cdot|-\rho_\mathrm{TF}*|\cdot|^{-1}$ can be written
as
\begin{equation}
  \label{eq:a0}
  \gamma_\mathrm{TF}\rho_\mathrm{TF}^{2/3}=V_Z
\end{equation}
(see, e.g., Gomb\'as \cite{Gombas1949}). This equation yields
immediately the upper bound
\begin{equation}
  \label{eq:a1}
  \rho_\mathrm{TF}(\bx)\leq (Z/\gamma_\mathrm{TF})^{3/2}|\bx|^{-3/2}.
\end{equation}
This bound allows us to prove the following $L^\infty$-bounds on 
potentials of exchange holes.
\begin{lemma}
  \label{l4} 
  $$\|L_{\rho_\mathrm{TF}}\|_\infty = O(Z).$$
\end{lemma}
\begin{proof}
  The function 
  \begin{equation}
    \label{eq:a3}
    \begin{split}
      f:\rz_+&\rightarrow \rz\\
      t&\mapsto \sqrt t\int_{|\by|<1/t}|\by|^{-1}|\by+(0,0,1)|^{-3/2}\rd \by
    \end{split}
  \end{equation}
  is obviously continuous on $(0,\infty)$. Moreover, $f(t)$ tends to a
  positive constant for $t\to0$ and to $0$ for $t\to\infty$. Thus,
  $\|f\|_\infty<\infty$.
 
  This allows us to obtain the desired estimate:
  \begin{equation}
    \label{eq:a4}
    L_{\rho_\mathrm{TF}}(\bx) \leq A_1(\bx) + A_2(\bx)
  \end{equation}
  where 
    \begin{multline}
      \label{eq:a5}
      A_1(\bx):=\int_{|y|\leq 1/Z} \frac{\rho_\mathrm{TF}(\bx+\by)}{|\by|} \rd
      \by 
      \leq \left(\frac Z{\gamma_\mathrm{TF}}\right)^{3/2}
      \int_{|y|\leq 1/Z} \frac{\rd \by}{|\by||\by+\bx|^{3/2}} \\
      = (Z/\gamma_\mathrm{TF})^{3/2} Z^{-1/2} f(|\bx|Z) \leq
      \|f\|_\infty\gamma_\mathrm{TF}^{-3/2} Z.
\end{multline}
and
\begin{equation}
  \label{eq:a6}
  A_2(\by):=\int_{\frac1Z\leq |\by| \leq R_{\rho_\mathrm{TF}}(\bx)}
  \frac{\rho_\mathrm{TF} (\bx+\by)}{|\by|} \rd \by \leq Z
  \int_{\frac1Z\leq |\by| \leq R_{\rho_\mathrm{TF}}(\bx)}
  \rho_\mathrm{TF}(\bx+\by)\rd \by
  \leq \frac Z2.
\end{equation}
These two estimate proof the claim.
\end{proof}

Lemma \ref{l4} allow us already to estimate the $N$ electron operator
$B_{c,N,Z}$ by the canonical one particle Brown-Ravenhall operator
whose nuclear charge is screened by the the Thomas-Fermi potential.
However, since we would like -- because of mere convenience -- to take
advantage of S\o rensen's result \cite{Sorensen2005}, we derive an
estimate on $L_{\rho_{\delta}}$ (where $\rho_\delta:=\rho_\mathrm{TF}*
g^2_{Z^{-\delta}}$), i.e., the exchange hole potential of the density
occurring in S\o rensen's proof.
\begin{lemma}
\label{l8}
$$\| L_{\rho_\delta}\|_\infty=O(Z).$$
\end{lemma}
\begin{proof}
  We proceed analogously to the proof of Lemma \ref{l4}:
  \begin{multline}
    \label{eq:a7}
    L_{\rho_\delta}(\bx) \leq \int_{|\by|\leq1/Z}
    \frac{\rho_\delta(\bx+\by)}{|\by|} \rd \by + \int_{1/Z\leq |\by|
      \leq R_{\rho_{\delta}}(\bx)}
    \frac{\rho_\delta (\bx+\by)}{|\by|} \rd \by\\
    \leq \int \rd \bz g^2_{Z^{-\delta}}(\bz)\int_{|\by|\leq1/Z}
    \frac{\rho_{\mathrm{TF}}(\bx-\bz+\by)}{|\by|} \rd \by +
    Z\int_{|\by|\leq R_{\rho_{\delta}}(\bx)}
    \rho_\delta(\bx+\by) \rd \by \\
    \leq \int \rd \bz g^2_{Z^{-\delta}}(\bz) A_1(\bx-\bz) + \frac Z2\leq k Z
  \end{multline}
  where we used the definition of the radius of the exchange hole from
  first line to the second line, the definition of $A_1$ in the next
  step, and in the last step the $L^\infty$-estimate \eqref{eq:a5} on
  $A_1$.
\end{proof}

\end{document}